\documentclass[10pt,twocolumn]{article}

\usepackage{latex8}  
\usepackage{times}
\usepackage{url}
\usepackage{comment}
\usepackage{epsfig}
\usepackage[compact]{titlesec}

\usepackage{fancyvrb}

\DefineShortVerb{\#}

\includecomment{comment}
\specialcomment{commentbf} {\begingroup\bf} {\endgroup}
\specialcomment{commentit} {\begingroup\it\footnotesize} {\endgroup}
\specialcomment{commentold} {\begingroup\it\footnotesize{\bf OLD:}} {\endgroup}
\excludecomment{comment}
\excludecomment{commentbf}
\excludecomment{commentit}
\excludecomment{commentold}

\title{
DPG:  A Cache-Efficient Accelerator for Sorting and for Join Operators
}
\author{
   Gene Cooperman\thanks{~$\,$This work was partially supported by
                the National Science Foundation under Grant CCR-0204113, and
                by the
                Institute for Complex Scientific Software
                (ICSS, http://www.icss.neu.edu/).},
 Xiaoqin Ma$^*$ and Viet Ha Nguyen$^*$
\\
Northeastern University\\
Boston, MA 02115, USA\\
\{gene,xqma,vietha\}@ccs.neu.edu
}

\begin{document}

\maketitle

\begin{abstract}
Retrieval of records on disk is well-known to be at the heart of many
database problems.  We show that the corresponding movement of records in main
memory has now become a severe bottleneck for many database
operations.  This is due to the stagnating latency of main memory,
even while CPU speed, main memory bandwidth, and disk speed all
continue to improve.  As a result, record movement has become the
dominant cost in main memory sorting.

We present a new algorithm for fast record retrieval, {\em
  distribute-probe-gather}, or DPG.  DPG has important applications
both in sorting and in joins.  Current main memory sorting algorithms
split their work into three phases: extraction of key-pointer pairs;
sorting of the key-pointer pairs; and copying of the original records
into the destination array according the sorted key-pointer pairs.
The copying in the last phase dominates today's sorting time.  Hence,
the use of DPG in the third phase provides an accelerator for existing
sorting algorithms.

DPG also provides two new join methods for foreign key joins: DPG-move
join and DPG-sort join.  The resulting join methods with DPG are
faster because DPG join is cache-efficient and at the same time DPG
join avoids the need for sorting or for hashing. The ideas
presented for foreign key join can also be extended to faster record
pair retrieval for spatial and temporal databases.

\end{abstract}

\section{Introduction}
\label{sec:intro}

Two important database operations are sorting and joins.  These
operations have three primary hardware-related costs: disk access, CPU
operation and main memory access.  The growth of main memory in
current computers implies that main memory databases become more
popular. For main memory databases, the bottleneck moves from disk to
main memory and the cost for disk access is not an issue anymore.

Further, the growing CPU-memory gap implies that CPU costs represent
an increasingly small portion of the total time.  This has been borne
out by several studies of
DBMs~\cite{Ailamaki99,BGB98,Patterson98,TLZT97}.  The impact of the
CPU-memory gap was popularized by the paper of Wulf and
McKee~\cite{WulfMcKee95} on the {\em memory wall}.

The diminishing role of the CPU in the total running time is partially
accounted for by increasing CPU speeds and greater on-chip functional
parallelism.  In part, it is also accounted for because most CPUs
today implement non-blocking caches and hardware prefetch in order to
overlap CPU execution with memory access~\cite{processor}.  Hence,
memory access becomes the bottleneck.  Therefore, we follow the
example of previous
researchers~\cite{Ailamaki99,Boncz99,Manegold99,Manegold00,Manegold02,ShatdalEtAl94}
in concentrating on main memory as the bottleneck.

At the heart of this memory bottleneck lies the {\em record retrieval
  problem}: the problem of copying records from a source array into a
  destination array according to a new ordering.  In a typical
  application, one will be given a source data file, a sequence of
  {\em record ids} (rids) for that data file, and a destination file.
  The task is to copy the source records into the destination file in
  the order specified by the sequence of rids.  Fast record retrieval
  is the key to faster sorting and faster joins.

A standard approach for data retrieval  accesses the records of the
source data file directly according to the sequence of the rids.  This
implies random access to the main memory.  This, for example, is what
was done in AlphaSort~\cite{AlphaSort} and
SuperScalarSort~\cite{SuperScalarSort}, the current record holders for
the Datamation sorting challenge~\cite{Datamation}.  However, the cost
of such random access has now become the dominant cost in main memory
sorting, since the CPU-memory gap has widened still further since the
original work on sorting.  (In fact, AlphaSort and SuperScalarSort
sort data on disk, but the size of the data file, 100~MB, is small
enough that the disk access consists solely of reading the source data
file from disk, and writing to a destination data file.)

Random access is harmful not just in disk resident databases, but also in main
memory resident databases.  Current DRAM technologies, such as DDR RAM and
Rambus RAM (RDRAM), extract a large latency penalty for any
non-sequential access to RAM.  This is because the memory chips are
divided into memory pages of several kilobytes, and there is a latency
penalty for switching to a new memory page~\cite{LostCircuits,DDR}.

Random access to RAM also harms performance in a second manner.
Random access incurs a heavy penalty when large cache blocks are used.
On a cache miss, the entire cache block is loaded into memory.  If the record size
is small compared to the cache block size, then there is a large
overhead  to load the entire cache block.  For example, for a cache
miss on a Pentium~4 with DDR-266 RAM, approximately 60~ns are spent
loading the cache block, and approximately 60~ns are spent waiting on
the latency of DDR RAM.  The trend is toward larger cache blocks.  The
128~byte L2 cache blocks of the Pentium~4 are four times larger than
those of the Pentium~III.  The IBM Power4 processor goes still further 
using 512~byte L3 cache blocks.

The solution to avoid these latency penalties in main memory is to
access main memory sequentially.  This is similar in spirit to the
way in which traditional databases strongly prefer to access disk
sequentially.  In analogy with operation on disk, two-pass algorithms
are a key for faster main memory performance.

The DPG-based sorting algorithms immediately yield faster sorting
algorithms.  Both AlphaSort and SuperScalarSort sort their data
essentially in three phases:  extraction of key-pointer pairs;
sorting of the key-pointer pairs; and copying of the original records
into the destination array according to the sorted key-pointer
pairs.  The last phase is essentially record retrieval.

A re-implementation of AlphaSort and SuperScalarSort on a IBM p690
Turbo shows that the record retrieval phase now dominates the running
time.  With the DPG record retrieval algorithm replacing the standard
record retrieval algorithm, we immediately produce a faster sorting
algorithm.  In the re-implementation of SuperScalarSort, the versions
with DPG as an accelerator are 27\% faster.


A direct consequence of faster record retrieval is faster main memory
sort-merge joins.  For example, in implementing sort-merge join using
SuperScalarSort, we find that the use of DPG sort instead of
SuperScalarSort results in a 27\% faster join algorithm..  The DPG
record retrieval phase in isolation is 48\% faster than traditional
record retrieval algorithm.

Finally, we apply DPG algorithm in the context of {\em foreign key
joins}.  In a foreign key join the join key is the same as the foreign
key.  we assume that one relation has an secondary index on the join
key.  Foreign key joins have the advantage that one need only
rearrange the records of one of the files.  This is in distinction to
sort-merge join and hash join, which both require to rearrange each of
the two files into a new file with join key values in sorted order.

Foreign key joins require less record retrieval because it is possible
to first extract {\em join triples}, ($k$, rid$_R$, rid$_F$), where
$R$ and $F$ are the two relations and $k$ is the join key value.
Assume that $R$ references a foreign key of~$F$.  To construct a join
triple, do file scan of~$R$, for each record of~$R$, its key~$k$ is
extracted.  The secondary B-tree index of the join key on~$F$ is then
used to derive the corresponding record id, rid$_F$, with the key
value~$k$. The standard index lookup is very expensive, we propose a
cache efficient B-tree batch lookup to generate the join triples.

Join triples reduce foreign key join to record pair retrieval.  The
join triples specify the record pairs, (rid$_R$, rid$_F$), to be
retrieved.  One can re-order the records of~$F$ to match the ordering
of rid$_R$ in the sequence of record pairs. Recall that, we do file
scan for~$R$, so rid$_R$s in the join triples are in sorted order.
Alternatively, one can sort the join triples according to rid$_F$, and
re-order the records of~$R$ to match the ordering of rid$_F$ in the
sorted rid pairs.  In either situation, there are fewer record
retrievals, and so foreign key join is faster than a general join.

The ideas of faster record retrieval can also be applied to the
general case of record pair retrieval.  Many algorithms for spatial
join~\cite{ArgeEtAl98,Patel96} and temporal join~\cite{Zhang02}
produce record pairs.  Unlike equijoin, there is no single search key,
and so record retrieval is more difficult.  The ideas of this paper
are described in terms of foreign key join.  However, it is even
simpler to translate the ideas into record pair retrieval, since an
initial join triple extraction is not required.

\section{Distribute-Probe-Gather}
\label{sec:DPG}

The {\em distribute-probe-gather algorithm} (DPG) is a record retrieval
algorithm.  Given a data file of records
and a sequence of record ids (rids) for the file, the goal is to copy
the records into a destination file with the property that
the ordering of records in the destination file corresponds to
the ordering of the rids in the given sequence.

For example, let the  source file, #R#, be the array of records with a
secondary \rm{B+} tree index on the attribute #A#. We want to
place the records of #R# in sorted order according to the values of~#A#. The
sequence of record ids, #S_rid#, at leaf nodes of the \rm{B+} tree is a list of record
ids in sorted order according to the value of A. Then, the destination file, #D#,
below, will contain the corresponding records in sorted order according to the
value of~#A#.
  
{\tt
\begin{quote}
for each i,
  D[i] = R[S\_rid[i]]
\end{quote}
}

In the case that the sequence of rids is a permutation of the rids for
the data file, the sequence of rids acts as a permutation vector.  The destination
file is then a permutation of the records in the input file.

Note, however, that the DPG algorithm is not limited to permutations
of data. In the case of join, one record from one relation may match more than one
records from another relation. In such cases portion of the original records must
be duplicated. The DPG algorithm also works for this case.

\subsection{\bf Algorithm}

The input for the algorithm is: a list of rids, #RID_LIST#, and a
data file of records, #INPUT#.  The DPG algorithm partitions the rids,
#RID_LIST#, into separate runs.  It also partitions the data records,
#INPUT#, into separate runs.  The algorithm makes two passes over the
record ids, #RID_LIST#, and two passes over the records of the data
file, #INPUT#.

The ideas are presented in the context of main memory databases.  We
assume that neither the sequence of rids, #RID_LIST#, nor the data
file, #INPUT#, fit in cache.  The DPG algorithm applies equally well
as an external data retrieval algorithm between disk and main memory.

The spirit of the DPG algorithm is: try to transform arbitrary memory
access patterns into sequential memory access patterns; where
arbitrary memory access patterns are unavoidable, we try to divide the
data into small partitions that fit into the cache.

For a sequential access pattern, it is easy to maintain a buffer in
cache.  In the DPG algorithm we need to read many streams
simultaneously, and maintain a buffer for each stream. Hence, we want
to keep the buffer as small as possible while maintaining reasonable
efficiency.  We define a buffer in cache to consist of two cache
blocks.  When a cache block is full, it is written back to main memory
and a new cache block is loaded from main memory into cache. On
Pentium~4, this is done automatically by the hardware prefetch
function unit if the access to main memory is sequential. On other
architectures without the hardware prefetch function unit, the
software instructions, $cflush$ and $prefetchnta$, are needed to
maintain the buffers.

\noindent
\paragraph{\bf Constraints:} The data records and rids are split into
runs of length~$L$.  The run length~$L$ is chosen based on two
constraints.  First, the cache must be able to simultaneously hold
both one run of data records of length~$L$ and one single buffer for the
corresponding run of rids.  (This constraint applies in the second
phase of DPG.)  Second, the cache must simultaneously be able to hold
a buffer for each run. (This constraint applies in the first phase and
in the third phase of DPG.)  These constraints are typical of the
constraints for two-pass algorithms, such as external sorting.

If the data file has $N$~records, then the data file is partitioned
into $N/L$ sets of consecutive records.  Assuming an rid consists of a
page id and offset on that page, the high order bits of the page
number can be used to efficiently identify the particular partition to
which the rid belongs.  This assumes that the number of pages in a
partition is a power of two, which can be satisfied by appropriate
choice of~$L$. For the sake of clarity, we assume the rids
values are in the rage of $0$ and $N$. 

\paragraph{\bf Three Phases:} There are three phases in the DPG
 algorithm. The three phases are also illustrated by pseudo-code in
 Figure ~\ref{fig:DPGalgo} and by the diagram of Figure~\ref{fig:DPG}.

\begin{enumerate}
\item{\bf Phase I.}  The first phase is the {\em Distribute}
phase. One {\em distributes} the rids of #RID_LIST# into appropriate
RID runs according to the values of the rids. The first RID run
contains the rid values in the range from $0$ to $L$, the second RID
run contains the rid values in the range from $L$ to $2L$, and so on.
Both the access to every RID run and the access to #RID_LIST# are
sequential. Therefore, one only needs to maintains a buffer in
cache for each RID run and a single buffer in cache for #RID_LIST#, At
the end, we form $N/L$ RID runs and each RID run is a permutation
vector. For example, the $i$-th RID run is a permutation vector in the
range from $(i-1)*L$ and $i*L$.

\noindent
\item{\bf Phase II.}  The second phase is the {\em Probe} phase.  In
this phase, we allocate a second, temporary data file, #INTERNAL#, in
main memory.  The temporary data file has the same size as the
original data file, #INPUT#, and is organized into the same number of
runs as the original data file. One then proceeds through each of the
runs of rids and each of the runs of #INPUT#.  The rids from the
$i$-th RID run are used to probe the $i$-th #INPUT# run and the
corresponding records are copied into the $i$-th #INTERNAL# run.  

At the end, the $i$-th #INTERNAL# run contains the same records as the
$i$-th #INPUT# run, but the order of records in the #INTERNAL# run is
organized according to the $i$-th RID run. Both the $i$-th RID run and the $i$-th
#INTERNAL# run are accessed sequentially. The $i$th #INPUT# run is
accessed randomly, but it can fit in cache. Hence, every time, one loads the
$i$-th #INPUT# run entirely into cache and maintains two buffers in cache: one
for the $i$th RID run and the other for the $i$-th #INTERNAL# run.

\noindent
\item{\bf Phase III.}  The third phase is the {\em Gather} phase.
This is an inverse of the {\em Distribute} phase and is similar to the
merge phase of external sorting.  In the {\em Distribute} phase, the
rids are distributed into runs.  As a result of the {\em Probe} phase,
the records in a given #INTERNAL# run are now in the same order as the
rids in the corresponding RID run created in Phase I.  Hence, it
suffices to gather (merge) the records from the #INTERNAL# runs in
exactly the same order as the order of the rids in #RID_LIST#.  More
precisely, if the $i$-th rid was distributed to the $j$-th RID run
during the {\em Distribute} phase, then at the $i$-th step of the {\em
Gather} phase, the next record from the $j$-th #INTERNAL# run is
copied to the destination array. One maintains several buffers in
cache: one single buffer for #RID_LIST#, one single buffer for the
destination array, and a buffer for each #INTERNAL# run.
\end{enumerate}

\subsection{\bf Example}
The DPG algorithm is presented more formally in pseudo-code in
Figure~\ref{fig:DPGalgo}.  The input of the algorithm is an array,
#RID#, of {\em rids} (record ids) and a data file, #INPUT_REC#, of
records.  It is desired to retrieve the records from #INPUT_REC# in
the order corresponding to #RID#.  The retrieved records are then
written to #OUTPUT_REC#.

The three phases of the DPG algorithm are illustrated by an example in
Figure~\ref{fig:DPG}.  The input in this example is the leaf nodes of
a secondary B+-tree index. The input contains a sequence of key-rid
pairs sorted according to the key values.  The output is a sequence of
records sorted according to the key values of the secondary index. The
output will either be stored again on disk or else pipelined to the
next stage. The ability of the DPG algorithm to take advantage of
pipelining is an important feature for sorting and joins.

The letters #a#, #b#, #c#, ... are used to indicate the sorted keys on
the leaf nodes of the index. So (#a#, 5) indicates that the record with the rid
value of~5 has a key value of #a#.  The first two rows are for
Phase~I, the next three rows for Phase~II, and the following
three rows for Phase~III.  The horizontal rectangle of the first
row represents a sequence of key-rid pairs sorted according to
the key values.  The second row represents the runs of rids into which
the first row is partitioned.  Similarly, the third row again
represents the runs of rids, but now as part of Phase~II.  The
fourth row is the partitioned runs of input records, and so
on. There are~12 elements and~3 runs in the example and each run contains
~4 elements. 

During {\bf Phase I}, one has a sequence of key-rid pairs sorted
according to the key values and will distribute them into appropriate
RID runs.  The first RID run will contain rid values from~0 to~3, the
second RID run will contain rid values from~4 to~7, and the third RID
run will contain rid values from~8 to~11. Upon reading the sequence of
pairs from the first row, one places the rids in their proper runs. For
example, the rid~5 goes to the second RID run, the rid~7 goes to
the second RID run, the rid~3 goes to the first RID run, the rid~8 
goes to the third RID run, and so on. The third row in the figure
presents the end of Phase I. In this
phase, we do sequential read on a single stream and sequential writes on multiple streams.

During {\bf Phase II}, one copies the original data file, #INPUT#, to
a temporary data file, #INTERNAL#.  The input to this phase is the
third row.  In the third row, each RID run has rid values that
correspond to one contiguous range of the #INPUT# file, an #INPUT#
run.  For example, for the first RID
run, we will load the first #INPUT# run into the cache.  The first
#INPUT# run consists of the first 4 contiguous records. The first rid
in the first RID run is~3 and $INPUT[3]$ is in the cache, so one finds
this record, #INPUT[3]#, in cache and copies it to the buffer
maintained in cache for the first #INTERNAL# run, and so on. At the
end of Phase~II, the records in the first #INPUT# run with key values in the sequence of i, l, f, c  are reordered as
records in the first #INTERNAL# run with key values in the sequence of c, f, i, l. The reordering is done according to a permutation specified by the first RID run, 3, 2, 0, 1.

During {\bf Phase III}, we will use the original rid sequence in the list of
key-rid pairs to {\em gather} (merge) records from all #INTERNAL# run. For
example, upon reading~5, we go to the second #INTERNAL# run to {\em gather}
the record; upon reading~7, we go to the second #INTERNAL# run to {\em gather}
the record; upon reading~3, we go to the first #INTERNAL# run to {\em gather}
the record; upon reading~8 we go to the third #INTERNAL# run to {\em gather}
the record, and so on. For this phase, we maintain a buffer for the key-rid list, a buffer for #OUTPUT# in cache and
a buffer for each #INTERNAL# run. All buffers are in cache.

\begin{figure}[htb]
\setbox0\vbox
{\small\begin{Verbatim}[commandchars=\\\{\},
  fontfamily=tt,
  commentchar=\%,codes={\catcode`$=3\catcode`^=7}]
% $ (for sake of emacs)
Let $L\gets$ (CACHE_SIZE/2)
Let $N\gets$ NUM_RECORDS
Let NUM_RUNS $\gets N/L$

\rm INPUT:  integer \tt RID[NUM_RIDS],
    \rm record \tt INPUT_REC[NUM_RECORDS];
\rm OUTPUT: record \tt OUTPUT_REC[NUM_RIDS];
\rm PARAMETERS: \rm integer \tt RID_RUN[NUM_RUNS][]\tt,
        \rm record \tt RECORD_RUN[NUM_RUNS][];

//Phase I:  {\it Distribute} RID array into runs
//          with each run of length $L$
% For $i$ = 1, ..., NUM_RUNS do
%   Initialize write buffer of RID_RUN[$i$]
For each rid, $r$, in RID do \{
 Set {\it run\_num} = $\lceil{}r/L\rceil$;
 Append $r$ to RID_RUN[{\it run\_num}]
\}

//Phase II:  {\it Probe} partitions of INPUT_REC
% (Conceptually partition INPUT_REC into partitions of length $L$ and
%  for each partition, $i$, of INPUT_REC, probe for each rid in RID_RUN[$i$])
For $i$ = 1, ..., NUM_RUNS do \{
%   Initialize read buffer for RID_RUN[$i$]
  Read into memory all records from
   INPUT_REC[$(i-1)\!*\!\!L\! +\! 1$] to INPUT_REC[$i\!*\!\!L$]
  Allocate memory for $L$ records
    to be stored in RECORD_RUN[i]
  For each rid, $r$, in RID_RUN[$i$]
    Append INPUT_REC[$r$] to RECORD_RUN[$i$]
  Write out RECORD_RUN[$i$] to disk
\}

//Phase III: {\it Gather} records from RECORD_RUN[]
//  into OUTPUT_REC in same order as RID[]
% For $i$ = 1, ..., NUM_RUNS do
%   Initialize read buffer of RECORD_RUN[$i$]
For each rid, $r$, in RID do \{
  Set {\it run\_num} = $\lceil{}r/L\rceil$;
  Read next record from RECORD_RUN[{\it run\_num}]
  Append record to OUTPUT_REC
\}
\end{Verbatim}
}
\centerline{\fbox{\box0}}
\caption{Distribute-Probe-Gather (DPG) Algorithm}\label{fig:DPGalgo}
\end{figure}


\begin{figure*}[!hbt]
\begin{center}
  \epsfig{file=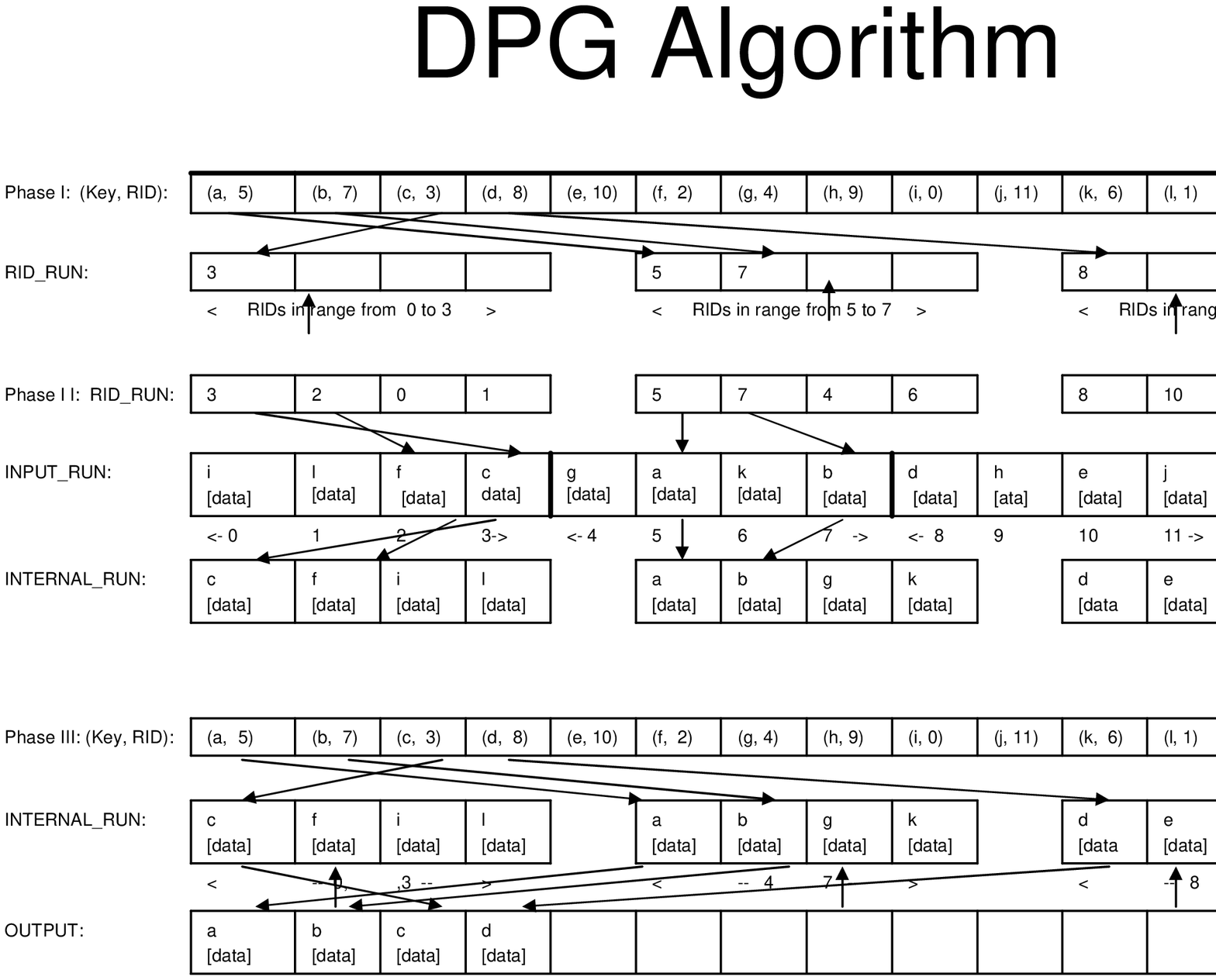,width=7.0in,clip=}
  \caption{Distribute-Probe-Gather (DPG)
    (also, see pseudo-code in Figure~\ref{fig:DPGalgo})}
  \label{fig:DPG}
\end{center}
\end{figure*}

\subsection{\bf Data Skew}
Implicit in the description of the DPG algorithm is that the input
sequence of rids is distributed uniformly among the set of all rids of
the input data file.  This is always the case when DPG is applied to retrieve records
after sorting key-pointer pairs.  In that situation, the
key-pointer pairs act as a permutation vector to permute the 
records in the input data file.

If the input sequence of rids is not uniformly distributed, then some
RID runs will be larger than other runs.  As a consequence, in
Phase~II, when the partition of the temporary data file (the partition
of #RECORD_RUN# in Figure~\ref{fig:DPGalgo}) may be larger than the size
of the cache.  If only a few of the partitions of the temporary data
file are larger than cache then the overall running time is not
greatly affected.

If there is a great deal of data skew and many of the temporary
partitions #RECORD_RUN# are larger than cache, then the $N/L$
partitions of the input sequence of rids must be chosen on some other
basis than the high order bits of the page number.  In such cases, one
can invoke the data skew handling techniques of DeWitt
et~al.~\cite{DeWitt92}.  Their solution, reformulated in our context,
is to sample the rids from the rid sequence.  The sampled set of rids
is then sorted, and partitions of the rids are chosen so as to evenly
partition the sampled set.

\section{Sorting}
\label{sec:sorting}

As discussed in the introduction, DPG acts as an accelerator for many
main memory sorting algorithms.  Recall that main memory sorts
typically proceed in three phases:
\begin{enumerate}
\item extraction of key-pointer;
\item sorting of the key-pointer pairs; and
\item copying of the original records
  into the destination array according to the sorted key-pointer
  pairs.
\end{enumerate}

AlphaSort~\cite{AlphaSort} and SuperScalarSort~\cite{SuperScalarSort}
are examples of this three-phase sorting paradigm.  Both sorting
algorithms can be considered as main memory sorting algorithms.

In principle, the sorting algorithms are single-pass disk-based
sorting algorithms.  Both sorting algorithms were introduced as an
answer to the Datamation Sorting Challenge~\cite{Datamation}.  The
Datamation challenge dictates that one is given one million records of
100~bytes.  Each record has a 10~byte key.  The keys are uniformly
distributed.  At the time of the Datamation Challenge, external
sorting algorithms were required.  On today's computers, the data file
of 100~MB easily fits in main memory.

Hence, the only disk-related portion of the Datamation Challenge is to
overlap disk~I/O with CPU operation.  Disk striping has the potential
to provide very fast disk~I/O.  This occurs because the disks are
accessed in parallel.  In this situation, main memory data retrieval
becomes the bottleneck.

The DPG algorithm pushes back this main memory bottleneck.  By using
DPG for data movement, the largest cost of main memory sorting is
reduced.  We have reimplemented the main memory portion of
SuperScalarSort, both with and without DPG.

\section {Join Methods with DPG}
\label{sec:joinAlgoDPG}

We use the ideas of DPG to present three new join algorithms: 1.c, 2 and~3.
Algorithms 1.a, 1.b and~4, will be included in the experimental
section~\ref{sec:joinExperiments} for completeness.
\begin{enumerate}
\item Sort-Merge Join
\begin{enumerate}
\item {Sort-Merge Join with AlphaSort (sort based on~\cite{AlphaSort})}
\item {Sort-Merge Join
    with SuperScalarSort (sort based on~\cite{SuperScalarSort})}
\item {Sort-Merge Join with DPG Sort}
  (sort based on DPG; see~\ref{sec:sortMergeJoin})
\end{enumerate}
\item {DPG-Sort Join} (see Section~\ref{sec:foreignKeyJoin})
\item {DPG-Move Join} (see Section~\ref{sec:foreignKeyJoin})
\item {Radix Join} (from~\cite{Manegold02})
\end{enumerate}

\subsection{Sort-Merge Join with DPG Sort}
\label{sec:sortMergeJoin}

The well-known Sort-Merge join was introduced by Blasgen and
Eswaran~\cite{Blasgen77}.  There are two steps in Sort-Merge joins:
sort two relations on the join key and scan the sorted relations to do
a merge on the join key.

Applying DPG sort at the first step provides a new faster sort-merge
join method, Sort-Merge join with DPG Sort. In
Sections~\ref{sec:sortMergeJoinExperiments}
and~\ref{sec:joinExperiments}, we experimentally compare different
versions of sort-merge join, according to the sorting methods used.
Specifically, we consider using DPG sort, AlphaSort and
SuperScalarSort for the sorting step.

\subsection{Foreign Key Join with DPG}

We next consider joins in which the join key is a foreign key, and it
has an index.
We denote by~$R$ a non-indexed relation.  We denote
by~$F$ an indexed relation.  The notation is motivated by the example
of a foreign key join.  In a foreign key join, the join key is the
same as the foreign key.  So, the join key is a set of attributes in
the relation~$R$ that refers to a foreign key from relation~$F$.

A {\em join triple} is a triple ($k$, rid$_R$, rid$_F$), such that $k$
is a key value, rid$_R$ is the rid of a record from~$R$ with key~$k$,
and rid$_F$ the rid of a record from~$F$ with key~$k$.

There are three steps in a foreign key join algorithm with DPG.  The
first step is to construct join triples.  The second step is to use
the join triples to copy one of the two relations into a temporary
file according to an order derived from the join triples.  The third
step is to join the temporary file with the remaining relation.

We describe the second and third steps initially in
Section~\ref{sec:foreignKeyJoin}.  We then return to the more
technical problem of efficiently constructing join triples in
Sections~\ref{sec:joinTriples} and~\ref{sec:indexLookup}.

\subsubsection{Two Foreign Key Join Methods with DPG}
\label{sec:foreignKeyJoin}

This section describes two DPG join algorithms.  It assumes that one
has already constructed the join triples.  Some algorithms for
constructing the join triples are described later in
Sections~\ref{sec:joinTriples} and~\ref{sec:indexLookup}.

Assume that one has generated the join triples ($k$, rid$_R$,
rid$_F$).  It is now possible to ignore the key~$k$, and deal directly
with the rid pairs (rid$_R$, rid$_F$).  We wish to satisfy one of two
goals:
\begin{enumerate}
\item\label{item:moveF} {\bf DPG-Move join:}  Move the records of~$F$ into
  a temporary file according to the ordering of the records of~$R$.
\item\label{item:moveR} {\bf DPG-Sort join:}  Move the records of~$R$ into
  a temporary file according to the ordering of the records of~$F$.
  So, the rid pairs (rid$_R$, rid$_F$) will be sorted according to
  rid$_F$.
\end{enumerate}

First consider DPG-Move join.  We will see how to generate the join
triples in the order of rid$_R$.  This is done by scanning the records
of the relation~$R$ in file order (in order of increasing rid$_R$).
Therefore the rid pairs can be used directly as part of a DPG
algorithm.  The second component of the pair, rid$_F$, is the sequence
of rids according to which we want to move the records of~$F$.  This
algorithm does not require any sorting or hashing.  Hence, we call
it DPG-Move join.

Next consider DPG-Sort join.  In this version, we first sort
the rid pairs (rid$_R$, rid$_F$) according to the order of rid$_F$.
This is done, for example, with SuperScalarSort and DPG.  The
algorithm then reduces to record movement in which we wish to move the
records of~$R$ according to the ordering of~rid$_F$ in the sorted
sequence (rid$_R$, rid$_F$).

Note that DPG-Move join is preferred when the relation~$F$ is smaller.
DPG-Sort join is preferred when the relation~$R$ is smaller.

\subsubsection{Construction of Join Triples}
\label{sec:joinTriples}

The simplest solution for a foreign key join is to
do a file scan of~$R$, and for each record of~$R$ to extract the join
key and do an index lookup in the index of~$F$.  One can then join the
record of~$R$ with the corresponding record of~$F$.  This involves
random access, and is economical only if the join produces
very few records.  In particular, this will be the case only if the
number of records of~$R$ is small.

A better solution is to do a file scan of~$R$, and to use the index
on~$F$ to create join triples.  To construct the join triples, one scans the
relation~$R$.  For each record of~$R$, one extracts the corresponding
rid and associated join key~$k$.  One then looks up the key~$k$ in the
index of~$F$.  The index lookup yields the final element of the
triple, rid$_F$.

\subsubsection{Batch Lookup in Indexes}
\label{sec:indexLookup}

Note that the join triple is constructed at the cost of a file scan
of~$R$ and an index lookup in the index of~$F$ for each record of~$R$.
Usually an index on a data file is much smaller than the full data
file.  If the index fits entirely in cache, then the index lookup will
be significantly cheaper than the file scan.

Unfortunately, the indexes in many main memory databases generally do
not fit in cache.  In such cases, the index lookup in the index of~$F$
will dominate the costs.  For example, in a B+~tree indexing
$N$~records, if an internal node has $m$~children, then $\log_m N$
nodes of the B+~tree must be accessed.  Each such access will be a
random access in main memory.  Most of the random accesses imply a
cache miss.  The cost of so many random accesses makes a naive index
lookup uneconomical.  Even if the index is a hash index, at least one
random access in memory will be required.

Luckily, it is possible to execute the index lookup faster than the
above analysis would indicate.  This is because the construction of
the join triples requires many index lookups, with no intervening
record accesses.  For purposes of join triple construction, {\em batch
lookup} of keys in an index suffices.  By batch lookup, we assume that
an array of join keys is first extracted by scanning the data file
of~$R$.  The batch lookup then produces an array of rids for~$F$
through the use of the index on~$F$.  We show that the index lookups
can be reorganized into a two-pass algorithm.  Two such two-pass
algorithms are demonstrated: one for B+ tree indexes, and one for hash
indexes.

Note that batch lookup of rids in an index can be substantially faster
than individual lookup.  Rao and Ross had previously discussed cache
conscious indexes for main memory~\cite{Rao99}.  There they present
CSS-trees, which have better cache behavior than either B+~trees or
hash indexes.  However, they only consider individual lookup of keys
in an index, one at a time.  In their scenario, a second key is not
looked up until the index lookup of the first key has been resolved.

\paragraph{\bf Batch Lookup in B+ Tree Indexes}

For simplicity, we describe the two-pass lookup for an enhanced
B+~tree.  The notion of {\em enhanced B+~tree} was introduced by Rao
and Ross~\cite{Rao99}.  The idea is that all slots of a B+~tree node
are used.  This is similar to compact B-Trees~\cite{CS83} or to the
ISAM method introduced by IBM~\cite{GR93}.  In the context of a
general B+~tree this can be accomplished by maintaining updates to the
B+~tree in a separate index, and then doing a batch update by
reorganizing the B+~tree.  For brevity, we will sometimes refer to
B+~tree, although in all cases, an enhanced B+~tree is intended.

Assume that we are executing a file scan of~$R$ with the purpose of
constructing the join triples.  We collect a sequence of keys from the
relation~$R$, and we wish to carry out a batch lookup of rids in the
enhanced B+-tree.  The rids will be used to retrieve records from the
foreign relation~$F$.  We assume that the B+~tree does not fit in
cache.  Our goal is a two-pass algorithm which will efficiently
accomplish batch lookup.

Assume that the B+~tree has $m$~entries per node.  Assume that the
B+~tree indexes $N$~records.  Then there are $\log_m N$ levels.  Assume
further that the first $(\log_m N)/2$ levels (the top half of the
B+~tree) fit in cache.  For $m\gg 2$, the top half of the tree has
approximately $m^{(\log_m N)/2}=\sqrt{N}$ slots.  For example, if
there are $N=10^9$ records, then there are 32,000 slots, which clearly
fit inside cache.

\begin{figure}[!hbt]
\begin{center}
  \epsfig{file=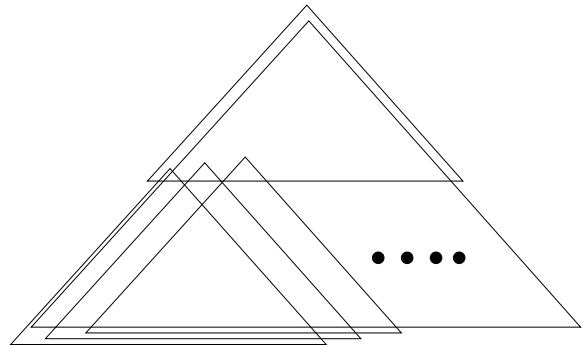,width=3.0in}
  \caption{B-Tree (Each smaller triangle represents a subtree that
  fits in cache.)}
  \label{fig:Btree}
\end{center}
\end{figure}

The strategy is a two-pass strategy.  In the first pass, one performs
a lookup of each key, but only using the top half of the B+~tree.
As shown in Figure~\ref{fig:Btree}, each leaf of the
subtree comprising this upper half can be considered as the root of a
second subtree comprising nodes in the lower half of the B+~tree.
Thus, at the end of this first pass, a key can be associated with a
leaf of the upper subtree, which is also a root of one of the lower subtrees.

After the first pass, one can associate each key with a subtree within
the lower half of the B+~tree.  So, in the second pass, one loads a
subtree from the lower half.  One then continues the lookup for all
keys associated with the root of the subtree in the lower half.
At the end, one then has a sequence of keys and rids.

If one wishes to have the keys in the same order as the order of the
original rids, then this can also be arranged.  In this case, one
extends the previous scenario to use the DPG algorithm.

The first phase of the DPG algorithm is to {\em distribute} the keys into
runs.  In the initial
lookup of a key in the first half, it was associated with a root of a
subtree in the first half.  The particular root of a subtree
identifies the run into which the key is copied.

The second phase of the DPG algorithm is to use the key to {\em probe} the
index, in order to find the rid.  One completes the lookup of all keys
in a run associated with a particular subtree of the B+~tree, before
proceeding to the next subtree in the lower half.  The resulting rids
are stored in a temporary partition or run.
This is exactly what was described earlier.

Finally, the third phase of the DPG algorithm {\em gathers} the rids
into a destination array in the same order as that of the original
keys.  Hence, we have completed a batch lookup of the keys, and
returned an array of rids in the same order as that of the original
keys.

\paragraph{\bf Batch Lookup in Hash Indexes}

Batch lookup of hash indexes also proceeds in two passes.  We assume
that the hash array of the hash index stores at each hash entry one
key-rid pair.  A second hash array associated with the index stores
pointers to overflow key-rid pairs that would have collided with an
occupied slot in the first hash array.

The two-pass lookup for hash indexes proceeds in a very simple manner.
We extract the sequence of keys from~$R$.  As the keys are extracted,
the hash values are computed.  Those key-hash value pairs are saved in
an array in an order corresponding to the rid order of~$R$.

It now suffices to apply the DPG algorithm.  The hash array is
partitioned into sets of $L$~hash slots.  In the distribute phase, the
hash value acts as an index into the hash array.  Hence, this becomes
the permutation vector of the DPG algorithm.  The key and hash value
are then written into separate runs according to the partition of the
hash array.  There is one run of key-hash values for each partition of
the hash array.

In the probe phase, a run of hash values is loaded into cache along
with the corresponding partition of the hash array.  As part of the
probe phase, the key and hash values from the run are used to look up
the the corresponding rid in the partition of the hash array.  The
rids are then saved in a temporary partition, in the same order as the
order in which the key-hash values of the original partition are stored.

Finally, in the gather phase, the ordering of the key-hash
value pairs are used to gather the rids from the temporary partition.
As in Section~\ref{sec:DPG}, the rids are gathered into a destination
array in an order corresponding the ordering of the original key-hash
value pairs.

\section{Experimental Evaluation}

\subsection{Sorting Comparisons}

\begin{figure}[!hbt]
\begin{center}
  \epsfig{file=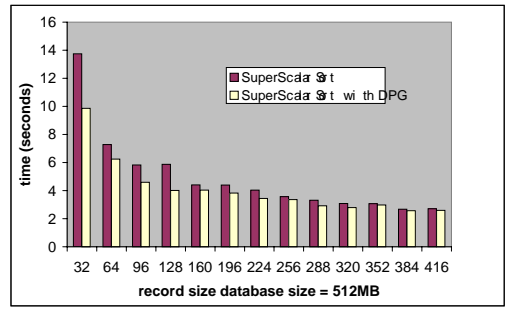,width=3.0in}
  \caption{Sorting Comparison}
  \label{fig:sorting}
\end{center}
\end{figure}

Figure~\ref{fig:sorting} demonstrates the acceleration achieved by
SuperScalarSort when DPG is used.  The results are demonstrated on the
IBM pSeries 690 Turbo.  The IBM~p690 has an L3~cache of size 128~MB.
Hence, in order to realistically demonstrate DPG, we were forced to
increase the size of the database.  We chose to implement
SuperScalarSort for a data file of size 512~MB.  The record size was
treated as a variable, to illustrate the influence of record size.  As
in the original Datamation Challenge, the key is 10~bytes.

Note that the additional speed of SuperScalarSort with DPG is most
pronounced for smaller record sizes.  For record sizes of 256~bytes
and higher, DPG provides only a small advantage.  This is because the
IBM p690 has an L3~cache block of size 512~bytes.  So, for record
sizes below 256~bytes, a cache miss incurs significant overhead
in loading a 512~byte cache block.

There was some variability in the results because the data was taken
on a time-shared, shared memory machine.  On the IBM~p690, the
L2~cache is shared among two CPUs, and the L3~cache is shared among
eight CPUs.  Our experiments were run on a single CPU.  Hence another
process on a neighboring CPU could consume some of our cache, thereby
affecting the timings.  The reported experimental results are the
averages of three runs each.

\subsection{Comparion of Sort-Merge Join using Different Sorting Algorithms}
\label{sec:sortMergeJoinExperiments}

\begin{figure}[!hbt]
\begin{center}
  \epsfig{file=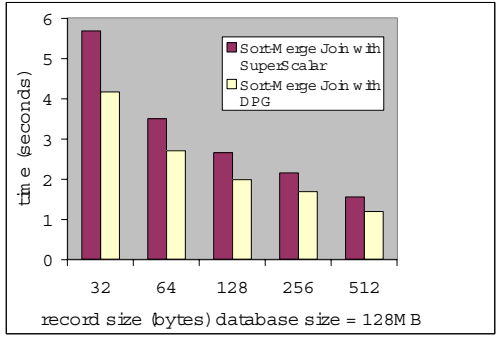,width=3.0in}
  \caption{Sort-Merge Join (IBM Power4 p690 Series)}
  \label{fig:sortMerge}
\end{center}
\end{figure}

We implement two sort-merge join methods: sort-merge join with DPG sort
and sort-merge join with SuperScalarSort.  As defined in section~\ref{sec:sortMergeJoin},
sort-merge join with DPG sort applies the DPG sort for the sorting
phase and sort-merge join with SuperScalarSort applies SuperScalarSort
for the sorting phase.

As expected, the acceleration in the speed of sort-merge join in
Figure~\ref{fig:sortMerge} follows the same pattern as that of
Figure~\ref{fig:sorting} for sorting. In this case, we illustrate our
results on a database of size 128~MB on the IBM p690.  The IBM p690
has an L2~cache of size 1.4~MB.  So, in this case, the L3~cache is acting
as the main memory, while L2~cache is acting as the ``cache''.

\subsection{Comparion of Six Different Join Methods}
\label{sec:joinExperiments}

\begin{figure}[!hbt]
\begin{center}
  \epsfig{file=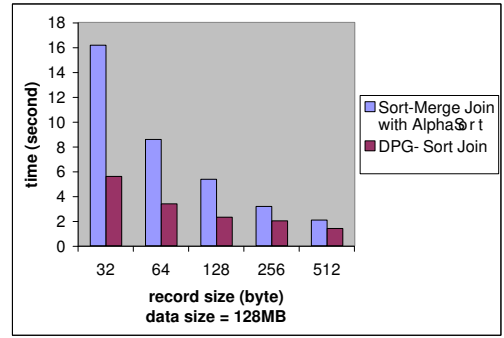,width=3.0in}
  \caption{Comparison of Joins (IBM Power4 pSeries 690 Turbo,
  exponential distribution of keys)}
  \label{fig:join_nonuniform_1}
\end{center}
\end{figure}

\begin{figure}[!hbt]
\begin{center}
  \epsfig{file=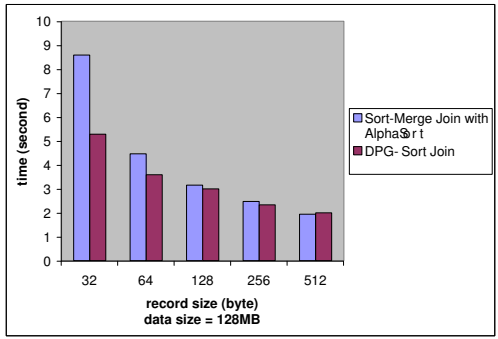,width=3.0in}
  \caption{Comparison of Joins (2.6~GHz Pentium 4 / DDR-266 RAM,
  exponential distribution of keys)}
  \label{fig:join_nonuniform_2}
\end{center}
\end{figure}

\begin{figure}[!hbt]
\begin{center}
  \epsfig{file=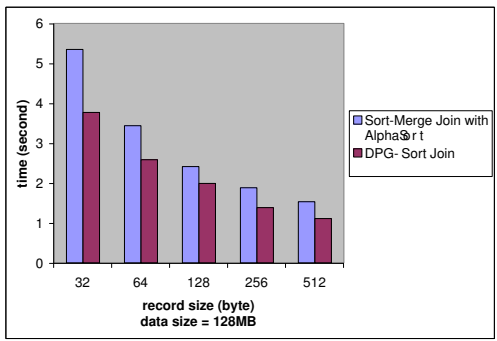,width=3.0in}
  \caption{Comparison of Joins (3.06~GHz Pentium 4 / Rambus PC-1200 RAM,
  exponential distribution of keys)}
  \label{fig:join_nonuniform_3}
\end{center}
\end{figure}

\begin{figure}[!hbt]
\begin{center}
  \epsfig{file=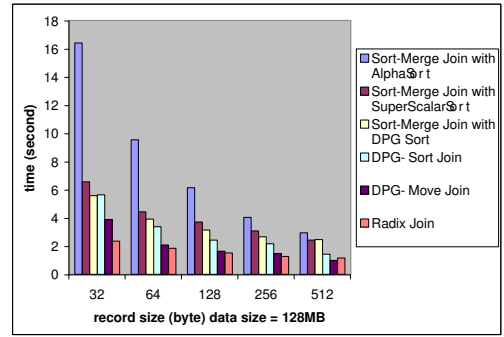,width=3.0in}
  \caption{Comparison of Joins (IBM Power4 pSeries 690 Turbo, no
  duplicate keys)}
  \label{fig:join_no_dup}
\end{center}
\end{figure}

\begin{figure}[!hbt]
\begin{center}
  \epsfig{file=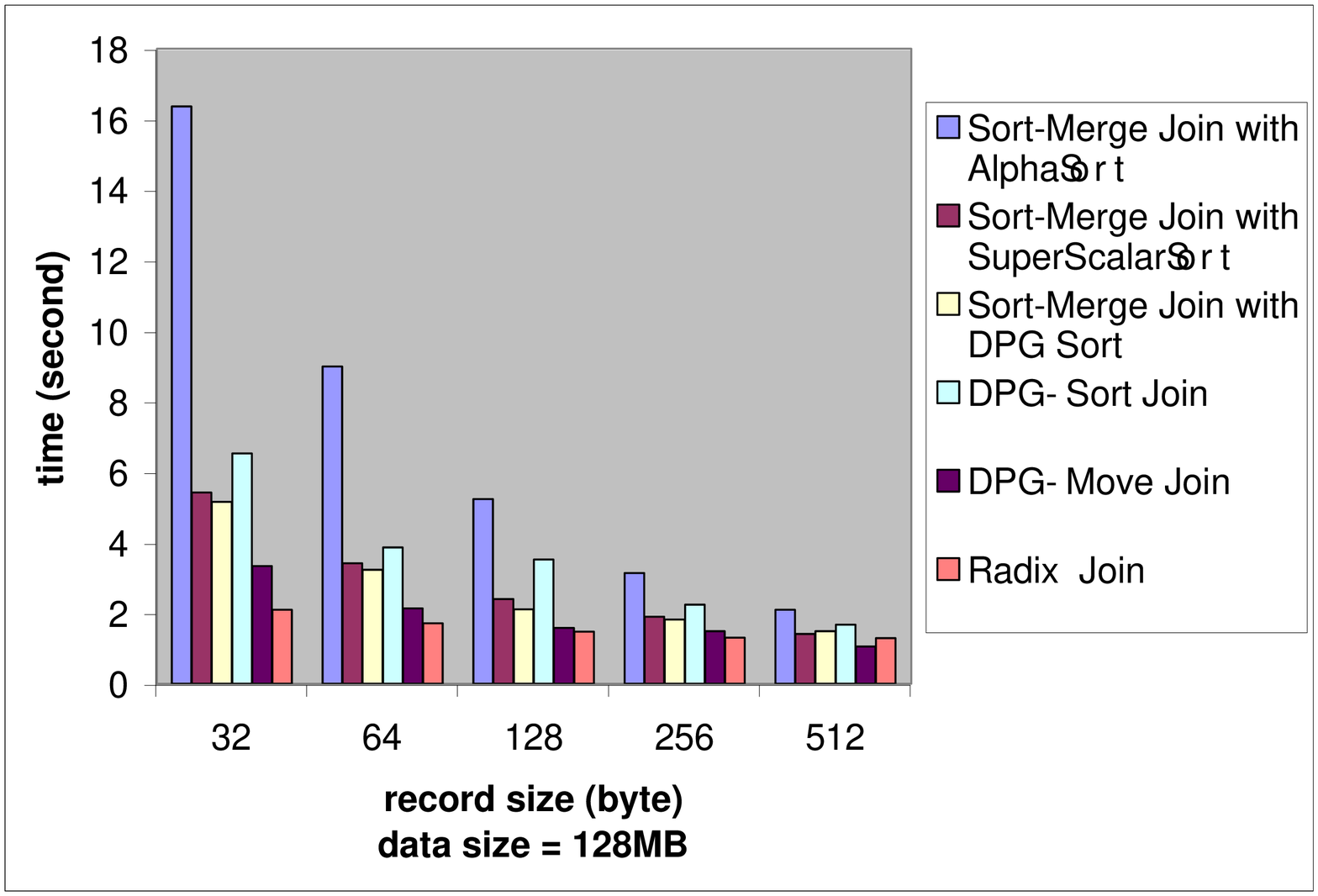,width=3.0in}
  \caption{Comparison of Joins (IBM Power4 pSeries 690 Turbo,
  duplicate keys)}
  \label{fig:join_dup_1}
\end{center}
\end{figure}

\begin{figure}[!hbt]
\begin{center}
  \epsfig{file=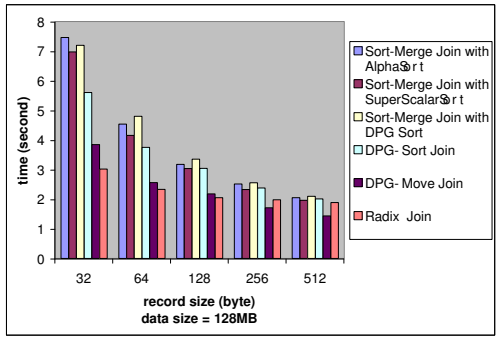,width=3.0in}
  \caption{Comparison of Joins (2.6~GHz Pentium 4 / DDR-266 RAM,
  duplicate keys)}
  \label{fig:join_dup_2}
\end{center}
\end{figure}

\begin{figure}[!hbt]
\begin{center}
  \epsfig{file=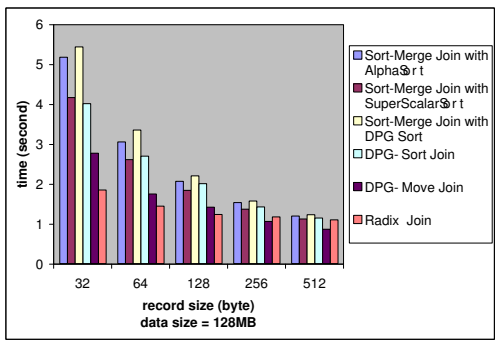,width=3.0in}
  \caption{Comparison of Joins (3.06~GHz Pentium 4 / Rambus PC-1200 RAM,
  duplicate keys)}
  \label{fig:join_dup_3}
\end{center}
\end{figure}

In Figures~\ref{fig:join_nonuniform_1} through~\ref{fig:join_dup_3},
we now experimentally compare the six join methods originally
presented at the begining of Section~\ref{sec:joinAlgoDPG}.  The three
join algorithms labelled Sort-Merge Join with DPG Sort, DPG-Sort Join,
and DPG-Move Join are all new.  The remaining algorithms, Sort-Merge
Join with AlphaSort, Sort-Merge Join with SuperScalarSort, and Radix
Join, are all based on sorting or join algorithms from the literature.

As explained earlier, we denote by
~$F$ an indexed relation and ~$R$ a relation that has foreign key on
the indexed attribute of ~$F$.
We consider both uniform and non-uniform distribution of foreign key values.

\paragraph{\bf The non-uniform distribution of join key values.}  Sort-merge join with
AlphaSort and DPG-Sort are the only two of the previously discussed join methods that operate
correctly for a non-uniform distribution of join key values.

In the following we use count bucket sort for bucket sort. The count bucket sort is as
follows:
\begin{enumerate}
\item count the number of elements destined for each bucket.
\item set bucket boundaries according to the statics computed and
distribute elements to buckets.
\end{enumerate}

AlphaSort works for non-uniformly distributed data, because it uses
quicksort to sort each run and uses replacement-selection to merge
the runs. DPG-Sort join works for non-uniformly distributed data too,
because we use count bucket sort to sort RIDs.  

DPG-Sort join sorts the RIDs according to the lowest~ $\log{N}$ bits
is enough, $N$ is the number of records. In the simulation, we assign
the RID values in the range $[1, ... , n]$, the lowest $\log{N}$ bits is
sufficient for sorting. For example, for $N=2^{20}$ sorting RIDs
according to the lowest 20 bits is enough. This could be done in two
steps with count bucket sort: first do count bucket sort according to
the lower 10 bits, then do count bucket sort according to the higher
10 bits.

Using the UNIX $random()$ and $exp()$ functions we generate an
exponential distribution of the data as $exp(c*(random()>>10))$ ($c$
is a constant and in our experiments we assign c with $-0.0000001$).  
A comparison of Sort-Merge join with AlphaSort and DPG-Sort join on
three different computer architectures is provided in the
figures~\ref{fig:join_nonuniform_1}, ~\ref{fig:join_nonuniform_2}, and
~\ref{fig:join_nonuniform_3}. The three different architectures are
the IBM Power4 Pseries 690 Turbo, the 3.06 GHz Pentium 4 with Rambus
PC-1200 RAM, and the 2.6 GHz Pentium 4 with DDR-266 RAM, From the
comparison, we can see that DPG-sort join is much faster than
Sort-Merge join with AlphaSort.

\paragraph{\bf The uniform distribution of join key values.}  First we will
show how all algorithms work for uniformly distributed join key values.
SuperScalarSort is a key-prefix-sort explained further in
section~\ref{sec:intro}.  It assumes that data is uniformly
distributed according to the highest ~7 bits of the key. This kind of
distribution can be applied to all DPG algorithms, because a unform
distribution of the join key values is the only constraint of
Sort-Merge join with DPG and DPG-move join.

We also implement the radix join of Manegold et al.~\cite{Manegold02}.
They describe a variation of hash join for main memory.  Radix join
also requires a uniform distribution of the join key values.
Otherwise, some of their partitions will be too large to fit into the
L2 cache and how to set the boundaries is unknown.

We produce uniformly distributed foreign key values using the UNIX
$random()$ function.  A comparison of the different join methods on
three different computer architectures as before is provided in the
figures. Figure~\ref{fig:join_no_dup} reflects data with no duplicate
join key values in the relation~$R$.  Figures~\ref{fig:join_dup_1},
~\ref{fig:join_dup_2}, and ~\ref{fig:join_dup_3}, show the same
information in which duplicate join key values are allowed.  From the
comparison, we can see that DPG-move join and radix join are the
fastest. DPG-move is better for large records and radix join is better
for small records.

\section{Conclusions}
The use of DPG in the sorting provides an accelerator for existing
sorting algorithms. Especially for the smaller record sizes, such as
32~bytes and 64~bytes, the performance improvements are really
impressive.

For the more common case of non-uniform distribution of join key values,
DPG-Sort join works better than sort-merge join with AlphaSort across
all the tested platforms. For smaller records sizes, such as, 32, 64,
DPG-Sort join is much better than sort-merge join with AlphaSort. More
impressively, on the newer platform, Rambus PC-1200 RAM, DPG-Sort works
better than on PC with DDR-266 RAM and even works better for larger
record sizes, for example 512~bytes.

For special case of uniform distribution of join key values, DPG-move join
and radix join are the best. The remaining DPG algorithms are also
competitive with older algorithms although with a smaller improvement.

\section{Future Work}
The DPG algorithm can be easily generalized to multiple passes.  This
can be useful when there is a very small cache in relation to the size
of main memory.  However, we do not encounter this scenario in our
current experiments.

\section{Acknowledgements}
We would like to thank Betty Salzberg and Donghui Zhang for extensive
conversations and insights into additional situations where
distribute-probe-gather can be beneficial.  We also gratefully
acknowledge the use of the support for the computations on the IBM
p690 by the Scientific Computing and Visualization (SCV) group at
Boston University.

\bibliographystyle{latex}

\end{document}